\documentclass[epj]{svjour}
\usepackage{default}

\usepackage[utf8x]{inputenc}
\usepackage[T1]{fontenc}

\usepackage{multirow}	
\usepackage{enumerate}	
\usepackage{Tabbing}	
\usepackage{relsize}	
\usepackage{textcomp}	

\usepackage{amsmath}
\usepackage{amssymb}
\usepackage{bm}		

\usepackage{hyperref}
\usepackage{url}
\usepackage{cite}	
\usepackage{mcite}	

\usepackage{dcolumn}	
\usepackage{longtable}
\usepackage{booktabs}	

\usepackage{graphicx}
\usepackage{rotating}	
\usepackage{float}	

\usepackage{color}

\newcommand{\average}[1]{\ensuremath{\left\langle#1\right\rangle}}
\newcommand{\bracket}[2]{\ensuremath{\left\langle#1 \vphantom{#2}\right| \left. #2 \vphantom{#1}\right\rangle}}
\newcommand{\matrixel}[3]{\ensuremath{\left\langle #1 \vphantom{#2#3} \right| #2 \left| #3 \vphantom{#1#2} \right\rangle}}

\newcommand{\mc}[3]{\multicolumn{#1}{#2}{#3}}

\newcommand{\Ang}{
\text{\AA{}}
}

\begin{document}
%

\title{Extended Hubbard model with renormalized Wannier wave functions in the correlated state III}
\subtitle{Statistically consistent Gutzwiller approximation and the metallization of atomic solid hydrogen}
\titlerunning{SGA and metallization of atomic solid hydrogen}

\author{Andrzej P. K\k{a}dzielawa\inst{1}\thanks{\emph{e-mail:} kadzielawa@th.if.uj.edu.pl} \and Jozef Spa\l{}ek\inst{1,2}\thanks{\emph{corresponding author:} ufspalek@if.uj.edu.pl}  \and Jan Kurzyk\inst{3} \and Włodzimierz Wójcik\inst{3}}

\institute{Marian Smoluchowski Institute of~Physics, Jagiellonian University, ul. Reymonta 4, PL-30-059 Krak\'{o}w, Poland \and Faculty of Physics and Applied Computer Science, AGH University of Science and Technology, ul. Reymonta 19,\\ PL-30-059~Krak\'{o}w, Poland \and Institute of Physics, Krak\'{o}w University of Technology, Podchor\k{a}\.{z}ych 1, PL-30-084~Krak\'{o}w, Poland}

\date{14 February 2013}

\abstract{
We extend our previous approach [J. Kurzyk, W. Wójcik, J. Spałek, Eur. Phys. J. B, \textbf{66}, 385(2008), and J. Spałek, J. Kurzyk, R. Podsiadły, W. Wójcik, Eur. Phys. J. B, \textbf{74}, 63(2010)] to modeling correlated electronic states and the metal-insulator transition by applying
the so-called \emph{statistically consistent Gutzwiller approximation} (SGA) to carry out self-consistent calculations of the renormalized single-particle Wannier functions
in the correlated state. The transition to the Mott-Hubbard insulating state at temperature $T=0$ is of weak
first order even if antiferromagnetism is disregarded. The magnitude of the introduced self-consistent magnetic correlation field is calculated and shown to lead to a small magnetic moment
in the magnetically uniform state. Realistic value of the applied magnetic field has a minor influence on the metallic-state characteristics near the Mott-Hubbard lcalization threshold.
The whole analysis has been carried out for an extended Hubbard model on a simple cubic (SC) lattice and the evolution of physical properties is analyzed as a function of the lattice
parameter for  the renormalized 1s-type Wannier functions. Quantum critical scaling of the selected physical properties is analyzed as a function of the lattice constant
$R\rightarrow R_c=4.1 a_0$, where $R_c$ is the critical value for metal-insulator transition and $a_0=0.53 \Ang$ is the Bohr radius. A critical pressure for metallization of solid
atomic hydrogen is estimated and is $\sim 10^2 GPa$.
\PACS{
      {71.30.+h}{Insulator-metal transitions} \and
      {71.27.+a}{Strongly correlated electron systems} \and
      {71.10.Fd}{Lattice fermion models} \and
      {62.50.-p}{High-pressure effects in solids and liquids}
     } 
} 
\maketitle
\section{Motivation}
\label{sec:mot}

 One of the important quantum-mechanical problems in both the solid-state \cite{MITorg,MIT,*Gebhard,Imada} and the optical-lattice \cite{Bloch} systems is the localization-delocalization
transitions of fer\-mionic states which is called \emph{the Mott} or \emph{Mott-Hubbard transition}. In electron systems it corresponds to the delocalization of atomic states
(usually of $3d$ or $4f$ type) and the formation of a Fermi liquid of heavy quasiparticles composed mainly of the transforming electrons, irrespectively of the fact
that there may be other valence electrons present also there \cite{Imada}. In the extreme version, the transition is driven solely by the interparticle interaction,
in which the presence of lattice plays only a secondary role. In that situation, the localization-delocalization transition is called \emph{the Wigner transition}
\cite{Wigner1,*Wigner2}. The description of these two transitions minimally provides a bridge between the atomic physics with the localized single-particle electron states and
the theory of fermionic quantum liquids with delocalized states, for which (quasi)momentum $\vec{p}=\hbar \vec{k}$ characterizes quasiparticle states. A principal dynamic
quantity driving the transition is the particle density (or interatomic distance for fixed number of particles, as is the case here).

A second impetus to the physics of these phenomena has been provided by the introduction by Anderson \cite{Anderson,*Seitz} and Hubbard \cite{Hubbard1} of second-quantization or
quantum-particle language, with the help of which the revised \emph{Mott-Hubbard transition} can be analyzed in terms of microscopic parameters. In the simplest, half-filled single-band model
the relevant microscopic parameter is the ratio $U/W$, where $U$ is the magnitude of the intraatomic (Hubbard) interaction and $W=2z|t|$ is the bare bandwidth (i.e., that for the uncorrelated
particles), with $z$ being the coordination number (i.e., the number of nearest neighbors) and $|t|$ is the magnitude of intersite transfer (hopping) of individual fermions. It is amazing
that a similar type of approach can be formulated for both fermions and the bosons, in the latter situation in the optical-lattice situation \cite{MIT,*Gebhard}.

The principal question is how to combine the Mott \cite{MITorg} and the Hubbard \cite{Hubbard2} aspects of this quantum phase transition in a purely electronic system. Once achived, the whole
description can be analyzed as a function of the lattice parameter (or interatomic distance, $R$). In the series of papers \cite{Kurzyk,Spalek} we have formulated such an approach starting
from the Gutzwiller-ansatz approximation (GA) for the extended Hubbard model, with a simultaneous readjustment of the single-particle Wannier functions in the correlated state.
The method provides, among others, a direct connection of the Mott criterion for localization/delocalization ($n_C^{{1}/{3}}a_B \approx 0.2$) with that of Hubbard ($U \approx W$).
As an extra bonus coming from such a formulation we obtain the quantum-critical behavior of single-particle wave-function size $\alpha^{-1}$ \cite{Spalek}, as well as the detailed
evolution of the correlated metallic state into the Mott-Hubbard insulator. One principal methodological advantage of the present formulation is that, in distinction to the LDA+U \cite{Anisimov1,Anisimov2}
or LDA+DMFT \cite{Kotliar,*Vollhardt} approaches our formulation avoids the problem of double counting of electron-electron interaction. However, unlike LDA+U or LDA+DMFT methods,
the present approach is still on the stage of modeling only the simplest (one-band) systems. Nonetheless, it may provide a formally proper starting point for more complex situations
such as many-band systems. Our formulations can certainly be also reformulated for Bose-Hubbard optical-lattice systems. It is unique in the sense of discussing of
quantum critical behavior of single-particle wave-function characteristics such as the inverse wave-function size (the inverse effective Bohr radius).

As a concrete application of our results we calculate the critical pressure for the metallization of solid atomic hydrogen with the electronic correlations included within
our renormalized mean-field theory.

The structure of the paper is as follows. In Sec.~\ref{sec:mod} we characterize briefly the method used in the paper and the modification of our previous approach \cite{Kurzyk,Spalek}. In Sec.~\ref{sec:res}
we analyze in detail the numerical results obtained with the help of the so-called \emph{Statistically Consistent Gutzwiller Approximation} (SGA). As a physical application, we
also provide there the estimate of the critical pressure for the solid atomic hydrogen metallization. Section~\ref{sec:con} contains an outlook with summary of main results and
a brief discussion of possible extensions.

\section{Model and Method Applied}
\label{sec:mod}

\subsection{Starting Hamiltonian}
\label{ssec:ham}

We start with the Extended Hubbard Hamiltonian for 1s hydrogenic-like system
 \begin{equation}
  \label{eq:HubbHam}
\begin{split}
  \mathcal{H} = &\epsilon_{a} \sum_i n_i + \sum_{i\neq j , \sigma} t_{ij} a^{\dagger}_{i \sigma} a^{}_{j \sigma} + U \sum_i n_{i\uparrow} n_{i\downarrow}\\
		 &+ \sum_{i < j} K_{ij} n_i n_j + \sum_{i < j} V_{ion-ion} \left( \vec{r}_j - \vec{r}_i \right),
\end{split}
 \end{equation}
where $t_ij$ is the hopping integral, $U$ the intraatomic interaction magnitude, $\epsilon_a$ the atomic energy per site, and $V_{ion-ion}$ corresponds to classical Coulomb interaction
between two $H^+$ ions, equal (in atomic units) to
\begin{equation}
 \label{eq:coulomb}
 V_{ion-ion} \left( \vec{r}_j - \vec{r}_i \right) = \frac{2}{\left| \vec{r}_j - \vec{r}_i \right|}
\end{equation}

By following \cite{RycerzPhD}, we introduce $N_e = \sum_i n_i$ - the total number of electrons, and $\delta n_i = n_i - 1$ as the deviation from neutral-atom configuration.
We can now rearrange the interatomic interaction in the following manner
\begin{equation}
 \label{eq:rearK}
 \begin{split}
  \sum\limits_{i < j} {K_{ij} n_i n_j } = &\sum\limits_{i < j} {K_{ij} } (n_i - 1)(n_j - 1) \\ 
 - &\sum\limits_{i<j} {K_{ij} + 2N_e \frac{1}{N}\sum\limits_{i < j} {K_{ij} } } \\
 = &\sum\limits_{i < j} {K_{ij} } \delta n_i \delta n_j + N_e \frac{1}{N}\sum\limits_{i < j} {K_{ij} } \\
 + &(N_e - N)\frac{1}{N}\sum\limits_{i < j} {K_{ij} }.
\end{split}
\end{equation}

By introducing now \emph{effective} atomic energy per site, i.e., containing both the atomic binding part $\epsilon_a$ and the ion-ion repulsion, we can write it down in the form
\begin{equation}
 \label{eq:Eeff}
 \epsilon_a^{eff} = \epsilon _a + \frac{1}{N}\sum\limits_{i < j} {\left( {K_{ij} + \frac{2}{{R_{ij} }}} \right)},
\end{equation}
where $R_{ij} \equiv \left| \vec{r}_j - \vec{r}_i \right|$. In effect, we can rewrite Hamiltonian \eqref{eq:HubbHam} in the following manner
 \begin{equation}
  \label{eq:ExtHubb}
\begin{split}
  \mathcal{H} = &\epsilon^{eff}_{a} \sum_i n_i + \sum_{i\neq j , \sigma} t_{ij} a^{\dagger}_{i \sigma} a^{}_{j \sigma} + U \sum_i n_{i\uparrow} n_{i\downarrow}\\
		 &+ \frac{1}{2} \sum_{i \neq j} K_{ij} \delta n_i \delta n_j.
\end{split}
 \end{equation}

We also add $( - \sum_{i,\sigma} \sigma \frac{1}{2} g \mu _B H_a n_{i \sigma})$ - a simple magnetic Zeeman term , where $g$ is the Land\'{e} factor, $\mu_B$ the Bohr magneton
and $H_a$ the external magnetic field. By introducing the reduced magnetic field $h \equiv \frac{1}{2} g \mu _B H_a$, we obtain our starting Hamiltonian
 \begin{equation}
  \label{eq:ExtHubbMag}
\begin{split}
  \mathcal{H} = &\epsilon^{eff}_{a} \sum_i n_i + \sum_{i\neq j , \sigma} t_{ij} a^{\dagger}_{i \sigma} a^{}_{j \sigma} + U \sum_i n_{i\uparrow} n_{i\downarrow}\\
		 &+ \frac{1}{2} \sum_{i \neq j} K_{ij} \delta n_i \delta n_j - h \sum_{i,\sigma}\sigma n_{i \sigma}.
\end{split}
 \end{equation}

This lattice Hamiltonian describing the system of 1s-type states in a solid contains the following microscopic parameters: $t_{ij}$, $U$, $K_{ij}$ and the band filling $n$.
Additionally, close to the metal-insulator boundary we can assume that $\average{\delta n_i \delta n_j}\simeq0$.

\subsection{Incorporation of single-particle wave function optimization}
\label{ssec:wav}

The microscopic parameters of \eqref{eq:ExtHubbMag} can be expressed via the single-particle Wannier functions in the following manner

\begin{equation}
 \label{meth:eq01.5}
 \begin{split}
 t_{ij} &= \matrixel{w_i}{H_1}{w_j},\\
 U &= \matrixel{w_i w_i}{\frac{\text{e}^2}{\left| \vec{r}_1 - \vec{r}_2 \right| }}{w_i w_i}, \\
 K_{ij} &= \matrixel{w_i w_j}{\frac{\text{e}^2}{\left| \vec{r}_1 - \vec{r}_2 \right| }}{w_i w_j}, \\
 \epsilon_a &= \matrixel{w_i}{H_1}{w_i}.
 \end{split}
\end{equation}

where $H_1$ is the Hamiltonian for a single particle in the system, and ${\text{e}^2}/{\left| \vec{r}_1 - \vec{r}_2 \right| }$ is interparticle interaction.
The numerical value is obtained by approximating first the Wannier functions $w_i \equiv w_i \left( \vec{r} \right) = w_i \left( \vec{r} - \vec{R}_i \right)$
by 1s Slater orbitals and those, in turn, by series of the Gaussian functions, i.e.,

 \begin{equation}
  \label{meth:eq01}
  \begin{split}
  w_{i} \left( \vec{r} \right) &= \beta \Psi_{i} \left( \vec{r} \right) - \gamma \sum_{j=1}^{z} \Psi_{j} \left( \vec{r} \right), \\
  \Psi_{i} \left( \vec{r} \right) &= \sqrt{\frac{\alpha^3}{\pi}} e^{- \alpha \left| \vec{r} - \vec{R}_i \right| } \\
		     &\approx \alpha ^{\frac{3}{2}} \sum_{a=1}^{p} B_a \left( \frac{2 \Gamma_a ^2}{\pi} \right) ^{\frac{3}{4}} e^{-\Gamma_a^2 \left| \vec{r} - \vec{R}_i \right|^2}.
  \end{split}
 \end{equation}

where the parameters $\beta$ and $\gamma$ are defined through Eqs.~(24) and (25) in part I\cite{Kurzyk} to make the basis normalized and orthogonal, i.e., $\bracket{w_i}{w_{j(i)}}=0$.
Parameters $B_a$ and
$\Gamma_a$ are derived by minimizing energy of single atom (Hamiltonian $\mathcal{H} \overset{a.u.}{=} -\bigtriangledown ^2 - 2{\left| \vec{r} - \vec{R}_i \right|^{-1}}$).
$p$~is the number of Gaussian functions used. Parameter $\alpha$ is found by minimizing the system energy of the trial correlated state (see below). The difference with
our previous approach \cite{Spalek} is that we include the statistical consistency conditions, as discussed next.

\subsection{Statistically-Consistent Gutzwiller Approximation (SGA)}
\label{ssec:SGA}

To obtain optimal value of the inverse size $\alpha$ given intraatomic distance $R$ we have to obtain the system energy. It was shown \cite{SGA} that the Gutzwiller Approximation (GA)
does not always provide the variational results consistent with those obtained from the corresponding self-consistent equations. To assure this consistency
we minimize the GA free-energy functional $\mathcal{F}$ supplemented with two additional molecular fields $\lambda_m$ and $\lambda_n$, coupled with $m$ and $n$
respectively:
 \begin{equation}
  \label{eq:SGA}
  \begin{split}
   \mathcal{F}^{(SGA)} = &- \frac{1}{\beta} \sum_{\vec{k}\sigma} \log \left( 1 + e^{-\beta E_{\vec{k}\sigma}^{(SGA)}}\right) \\
			 &+ \Lambda \left( \lambda_n n + \lambda_m m + U d^2 + \mu n \right),
  \end{split}
 \end{equation}

where the trial eigenvalues $E_{\vec{k} \sigma}^{(SGA)}$ are:

\begin{equation}
\label{eq:E_k}
  \begin{split}
E_{\vec{k}\sigma}^{(SGA)} &\equiv E_{\vec{k}\sigma} - \sigma \lambda_m - \lambda_n \\
&= q_\sigma \varepsilon_{\vec{k}} - \sigma \left( h + \lambda_m \right) - \left( \mu + \lambda_n \right),
  \end{split}
\end{equation}

where $d^2 = \average{n_{i\uparrow} n_{i \downarrow}}$ , and
\begin{equation}
\begin{split}
q_\sigma = \left( \sqrt{ \left( n_\sigma - d^2 \right) \left( 1- n_\sigma - n_{\overline{\sigma}} + d^2 \right) } \right. \\
+ \left. d \sqrt{ n_{\overline{\sigma}} - d^2} \right) ^2 
 / n_\sigma \left( 1 - n_\sigma \right)
\end{split}
\end{equation}
 
is the band narrowing renormalization factor and $\varepsilon_{\vec{k}}$ is dispersion relation for bare particles (here taken for simple cubic structure, 
$\epsilon_{\vec{k}} = 2 t\left( \cos k_x  + \cos k_y  + \cos k_z  \right)$). The eigenvalues $E_{\vec{k} \sigma}$ are obtained by Fourier transform of the effective
GA Hamiltonian

\begin{equation}
  \label{eq:FourierHamiltonian}
\begin{split}
  \mathcal{H} = \epsilon^{eff}_{a} \sum_{i \sigma} n_{i \sigma} &+ \sum_{i j \sigma} t_{ij} q_\sigma a^{\dagger}_{i \sigma} a^{}_{j \sigma} + \\
		&+ \Lambda U d^2 - \mu \sum_{i \sigma} n_{i \sigma},
\end{split}
 \end{equation}

additionally supplemented with the Lagrange-multiplier constrains

\begin{equation}
 -\lambda_m \sum_i \left( m_i - m \right) - \lambda_n  \sum_i \left( n_i - n\right),
\end{equation}

where $m_i \equiv n_{i\uparrow} - n_{i\downarrow}$, $m\equiv \average{m_i}$, $n_i \equiv n_{i\uparrow} + n_{i\downarrow}$, and $n \equiv \average{n_i}$.

The operator $\mathcal{K} \equiv \mathcal{H}- \sum_i ( \lambda_m m_i  + \lambda_n  n_i ) + \Lambda ( \lambda_m m  + \lambda_n  n )$ plays the role of the effective Hamiltonian,
in which the mean fields ($m$, $d^2$) and the Lagrange multipliers ($\lambda_m$, $\lambda_n$), as well as $\mu$, are all determined variationally, in addition to the wave function
parameter $\alpha$. In the next section we analyze in detail the physical results obtained by minimizing though that procedure the functional \eqref{eq:SGA}.

\subsection{Overview of numerical methods}
\label{ssec:num}

Numerical analysis was carried out with the help of method different from that used in \cite{Kurzyk,Spalek} by introducing the lower-level minimization for each single-particle
basis optimization step. In other words - for each and every step of minimizing energy with respect to the reverse function size $\alpha$, where
\emph{Golden Section Search} \cite{Press:2007:NRE:1403886} was empirically proven to be the most efficient, there is introduced a new minimization of functional $\mathcal{F}^{(SGA)}$ with
respect to the double occupancy $d$, the magnetization $m$, the chemical potential $\mu$ and the molecular fields $\lambda_m$ and $\lambda_n$. The latter procedure was carried using
\emph{GSL - GNU Scientific Library} (\url{http://www.gnu.org/software/gsl/}), with the order of magnitude of zero-precision $10^{-8}$ and all the following up calculations
with the double precision.

Due to new minimization step and new parameter (external magnetic field $H_a$), the numerical complexity increases by two orders of magnitude, and it enforces new optimization.
To decrease the computing time we chose the basis of three Gaussians ($p=3$ in \eqref{meth:eq01}) instead of seven (as in \cite{Kurzyk,Spalek}). After comparing the results for
constant external magnetic field $H_a$ for $p=3$ and $p=7$ we observe no qualitative change of behavior. Below we discuss the basic physical properties obtained within SGA
and just discussed numerical procedure, as well as compare them with those obtained previously without the statistical consistency \cite{Kurzyk,Spalek}.

\section{Results and discussion}
\label{sec:res}

\subsection{Ground-state properties}
\label{ssec:prop}

\begin{figure}
\centering
\includegraphics[width=0.99\linewidth]{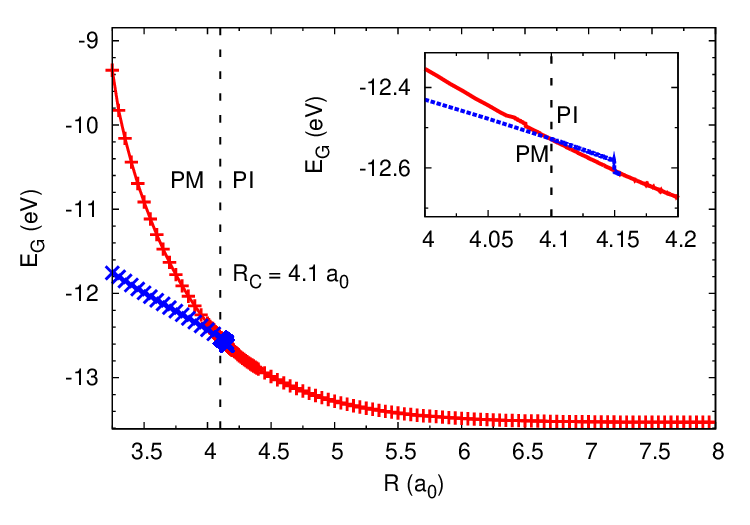}
\caption{Ground state energy (per atom) of the metallic state $(PM, \times)$ for $R<R_c$ and the insulating $(PI,+)$ for $R\geqslant R_c$, as a function of interatomic distance $R$
(in units of Bohr radius $a_0$). Inset: detailed representation of the first-order $PM \rightarrow PI$ transition near $R=R_c \approx 4.1 a_0$. The upper curve for $R<R_c$ represents
the energy of the unstable PI state. Note that as $E_G > -1 Ry$, the lattice can only be stabilized by the external pressure (see sec. \ref{ssec:Hmet}).}
\label{fig:met-ins}
\end{figure}

The calculated ground state energy $E_G$ as a function of interatomic distance $R$ (lattice parameter) is shown in Fig. \ref{fig:met-ins}. In the inset a detailed dependence of $E_G$
in the transition regime $R\approx R_c = 4.1 a_0$ is displayed. For the sake of comparison, the energy of $PI$ state (for $H_a=0$) has been also shown for $R<R_c$, where this phase
is not stable. The principal difference with out previous Gutzwiller ansatz (GA) results is that in the present (SGA) approach the transition is weakly discontinuous, as one can see
explicitly from the circumstance that ${dE_G}/{dR}$, which is proportional to the internal pressure, is discontinuous.

\begin{figure}
\centering
\includegraphics[width=0.99\linewidth]{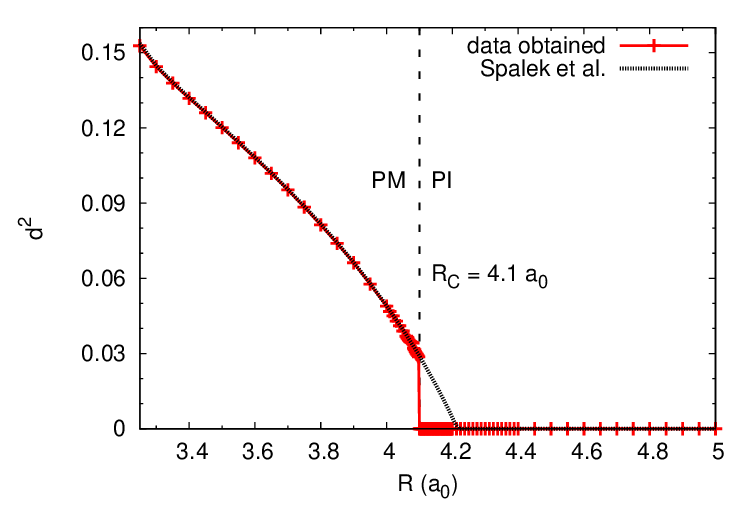}
\caption{Double occupancy probability $d^2 = \average{n_{i\uparrow} n_{i \downarrow}}$ versus $R$. Note a weak discontinuous jump to zero at $R=R_c \approx 4.1 a_0$, as compared
to the continuous evolution for $PM \rightarrow PI$ of the Gutzwiller approximation obtained previously \protect\cite{Kurzyk,Spalek}.}
\label{fig:dpow2}
\end{figure}

In Fig. \ref{fig:dpow2} we plot the double occupancy probability $d^2 = \average{n_{i\uparrow} n_{i \downarrow}}$ versus $R$ and again see a weak discontinuity at $R=R_c$. The present
SGA results are compared with the previous GA results of Spałek \emph{et al.} \cite{Spalek}.

\begin{figure}
\centering
\includegraphics[width=0.99\linewidth]{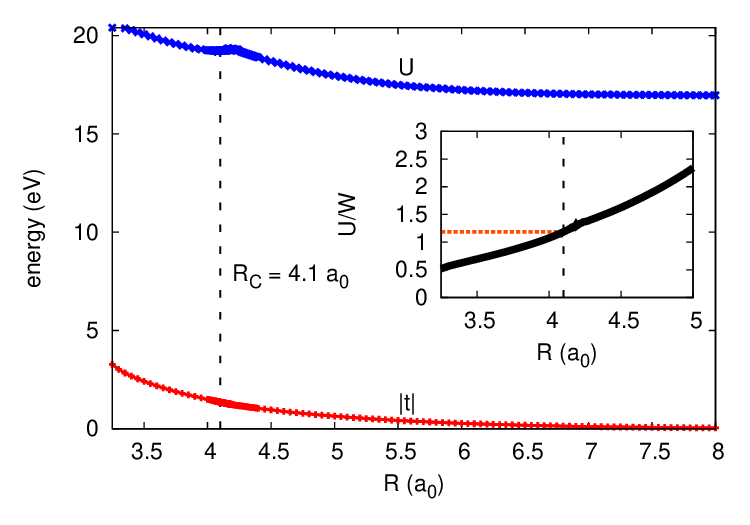}
\caption{Microscopic parameters: hopping integral $|t|$ and the Hubbard interaction parameter $U$, both as a function of $R$. Inset: $U/W$ ratio vs R. The $U/W$ ratio for $R=R_c$ is $\left(U/W\right)_C \simeq 1.18$
. The vertical dashed line marks the $PM \rightarrow PI$ transition point. In the large-$R$ limit $U$ reaches the atomic value $U_{at}=\left(5/4\right) Ry$.}
\label{fig:UtUW}
\end{figure}

\begin{figure}[H]
\centering
\includegraphics[width=0.99\linewidth]{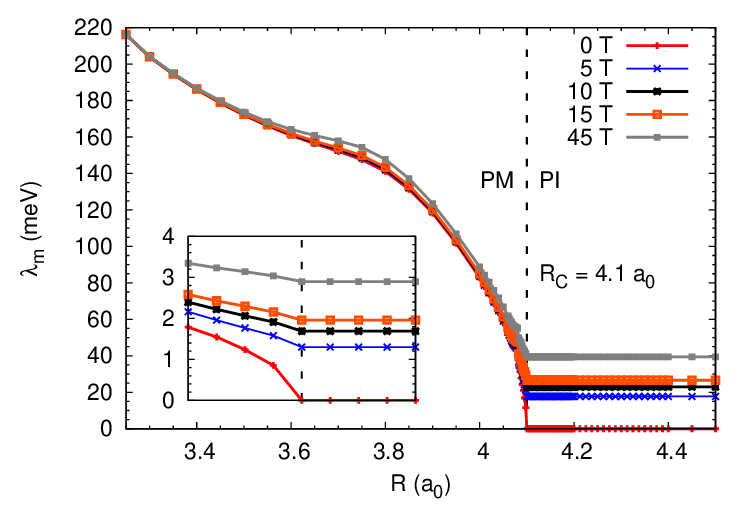}
\caption{Correlation-induced effective magnetic field $\lambda_m$ vs. $R$ for selected values of applied magnetic field $H_a$. For $H_a=0$, $\lambda_m$ vanishes at $PM - PI$ boundary (see inset for details).
The presence of the applied field triggers $\lambda_m \neq 0$ in the insulating state.
}
\label{fig:lmb_m}
\end{figure}

The $R$-dependent evolution of the microscopic parameters are shown in Figs.~\ref{fig:UtUW} and \ref{fig:lmb_m}. In Fig. \ref{fig:UtUW} we plot the values of $U$ and the nearest-neighbor hopping
magnitude $|t|$, whereas in the inset the $U/W$ ratio, with $W=2 z |t|$ being the bare bandwidth, is shown in the regime $R \approx R_c$. Note that at $R_c=4.1 a_0$ the $U/W$ is of
the order of unity, in accordance with the results obtained when $U/W$ is treated as a free parameter \cite{Hubbard2}. The atomic limit is reached effectively when $|t|\approx 0$, i.e.
for $R \sim 6.5 a_0$. The value of $U$ in that limit is $U=\left( {5}/{4} \right) Ry$. Amazingly, one should also note that at the transition $R_c^{-1}\alpha^{-1} =
n_c^{{1}/{3}} a_B \sim 0.25$, in accordance with the original criterion due to Mott \cite{MIT,*Gebhard} for the Mott localization ($n_C = {1}/{R_c ^3}$ is the particle density at
$R=R_c$ and $a_B=\alpha^{-1}$ in the effective Bohr radius at that point). This approach does allow to relate directly the Mott and the Hubbard criteria for localization
of fermions. This was achieved by readjusting the Wannier functions, determining $t$ and $U$ parameters in the correlated state.

In Fig. \ref{fig:lmb_m} we exhibit the evolution of the consistency field $\lambda_m$, which plays the role of the self-consistent (correlation) field, also for selected nonzero
applied magnetic field. One should note that the lowermost curve corresponds to the case with $H_a=0$. The nonzero value of the effective field in the metallic phase will introduce
a small but nonzero value of the spin magnetic moment in the uniform (weakly magnetic) phase as discussed below. The situation is highlighted in the inset to this figure. However, one
should emphasize that the field $\lambda_m$ is dependent on the phase discussed and in the antiferromagnetic state it takes the form of a staggered field \cite{Korbel}. The latter
phase will not be analyzed in detail here, as the first and the foremost aim of this paper is to underline the first-order nature of the $M \rightarrow I$ transition if the statistical
consistency conditions are properly taken into account, as well as the quantum scaling of the single-particle atomic wave function inverse size $\alpha ^{-1}$. The corresponding
behavior of the renormalized but less pronounced (cf. \cite{Spalek}), so we will not discuss it in detail here.

\begin{figure}
\centering
\includegraphics[width=0.99\linewidth]{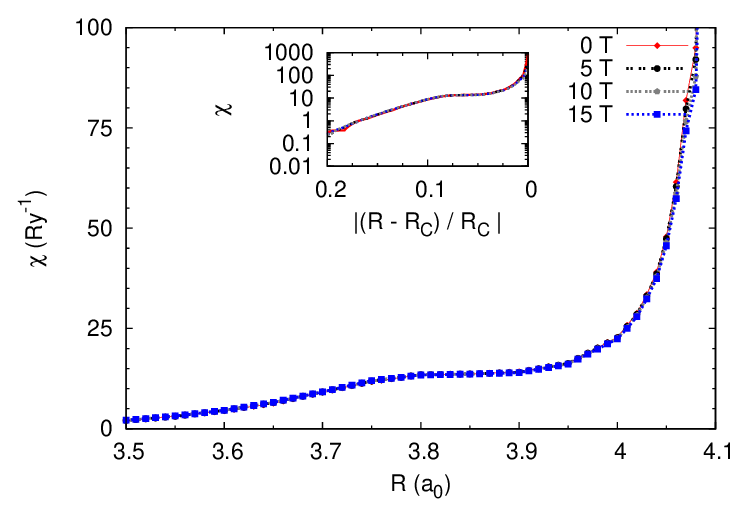}
\caption{Differential static magnetic susceptibility vs. $R$ for selected values of the applied magnetic field. The $\chi$ divergence at $R_c=4.1 a_0$ accompanies the $PM-PI$ transition
and is associated with localization of the itinerant electrons when $R \rightarrow R_c$. Overall $\chi$ behavior in the metallic state does not depend much on the value of $H_a$.
Inset: double logarithmic plot $\chi (R)$ showing absence of any simple exponential type of scaling.}
\label{fig:sus}
\end{figure}

\begin{figure}
\centering
 \includegraphics[width=0.99\linewidth]{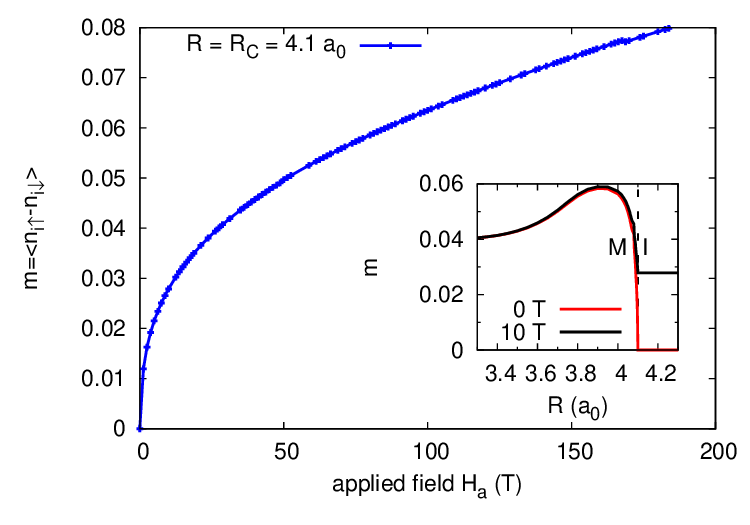}
\caption{Magnetic moment $m=\left< n_{i\text{\textuparrow}} - n_{i\text{\textdownarrow}} \right>$ as a function of the applied magnetic field at the critical interatomic
distance $R_c=4.1 a_0$. Inset: a residual uniform moment in the metallic state vs. $R$ near $R_c$, induced by the correlation field when it is assumed as spatially uniform.}
\label{fig:m_Rconst}
\end{figure}

To visualize directly the type of magnetism accompanying the metal-insulator transition (MIT), we have plotted in Fig. \ref{fig:sus} the zero-field magnetic susceptibility versus $R$
(the inset illustrates an overall behavior). The principal feature of $\chi$ quasi-divergence for $R \rightarrow R_c$ is roughly independent of the applied field magnitude specified.
It does not follow exactly the Brinkman-Rice dependence \cite{Brinkman} as here the unrenormalized $\left( \chi_0 \right)$ value is strongly $R$ dependent as well, so $\chi$
cannot be parametrized by $U/W$ ratio only. For the sake of completeness we
plot in Fig. \ref{fig:m_Rconst} the magnetization curve at the localization threshold. The dependence roughly emulates the Brillouin curve for the localized particles of spin ${1}/{2}$.
However, one has to take into account a nonlinear increase of the molecular field for nonzero $H_a$ value. (cf. Fig. \ref{fig:lmb_m}). In the inset to this figure we provide
the uniform magnetic-moment magnitude as a function of $R$ if a uniform $\lambda_m \neq 0$ is assumed. In the metallic state the spontaneous value of spin polarization
is very small and practically field independent.

\begin{table*}
\centering
 \caption{\label{tab:01} Computed quantities in SGA (except additional results for ground-state energy added, calculated in GA), all as a function of the lattice parameter, for
the simple cubic structure. For their definition, see main text. The values, if not specified explicitly, are in rydbergs ($Ry$). The $\chi(0)$ value for $R\geqslant R_c$ is infinite.}
\resizebox{\textwidth}{!}{
\begin{tabular}{|c||c|c||c|c||c||c|c||c|c}
\hline
\hline
\mc{1}{|c||}{$R (a_0)$}	& \mc{1}{c|}{$E_G^{SGA}$}	& \mc{1}{c||}{$E_G^{GA}$}	& \mc{1}{c|}{$t$}	& \mc{1}{c|}{$U$}& \mc{1}{c|}{$\alpha^{-1} (a_0)$}	& \mc{1}{c|}{$d^2$}	& \mc{1}{c|}{$\lambda_m$}	& \mc{1}{c|}{$\chi (Ry^{-1})$} & \mc{1}{c||}{$q^{-1}$}\\
\hline
\mc{1}{|c||}{3.25}	& \mc{1}{r|}{-0.8640}		& \mc{1}{r||}{-0.8644}		& \mc{1}{r|}{-0.2409}	& \mc{1}{r|}{1.4996}	& \mc{1}{r|}{0.9474}		& \mc{1}{r|}{0.152774}	& \mc{1}{r|}{0.015884}		& \mc{1}{r|}{0.1809} & \mc{1}{r||}{1.17728}\\
\mc{1}{|c||}{3.50}	& \mc{1}{r|}{-0.8814}		& \mc{1}{r||}{-0.8816}		& \mc{1}{r|}{-0.1773}	& \mc{1}{r|}{1.4749}	& \mc{1}{r|}{0.9220}		& \mc{1}{r|}{0.120128}	& \mc{1}{r|}{0.012641}		& \mc{1}{r|}{2.0598} & \mc{1}{r||}{1.36818}\\
\mc{1}{|c||}{4.00}	& \mc{1}{r|}{-0.9136}		& \mc{1}{r||}{-0.9136}		& \mc{1}{r|}{-0.1098}	& \mc{1}{r|}{1.4152}	& \mc{1}{r|}{0.9200}		& \mc{1}{r|}{0.048886}	& \mc{1}{r|}{0.006084}		& \mc{1}{r|}{22.6577} & \mc{1}{r||}{2.82781}\\
\mc{1}{|c||}{4.05}	& \mc{1}{r|}{-0.9171}		& \mc{1}{r||}{-0.9171}		& \mc{1}{r|}{-0.1046}	& \mc{1}{r|}{1.4139}	& \mc{1}{r|}{0.9175}		& \mc{1}{r|}{0.038973}	& \mc{1}{r|}{0.004256}		& \mc{1}{r|}{47.3562} & \mc{1}{r||}{3.47235}\\
\mc{1}{|c||}{4.09}	& \mc{1}{r|}{-0.9200}		& \mc{1}{r||}{$ $}		& \mc{1}{r|}{-0.1005}	& \mc{1}{r|}{1.4140}	& \mc{1}{r|}{0.9147}		& \mc{1}{r|}{0.030193}	& \mc{1}{r|}{0.027281}		& \mc{1}{r|}{253.7567} & \mc{1}{r||}{4.40375}\\
\hline
\mc{1}{|c||}{4.10}	& \mc{1}{r|}{-0.9209}		& \mc{1}{r||}{-0.9207}		& \mc{1}{r|}{-0.0995}	& \mc{1}{r|}{1.4143}	& \mc{1}{r|}{0.9138}		& \mc{1}{r|}{0.000000}	& \mc{1}{r|}{0.000000}		& \mc{1}{r|}{$\infty$} & \mc{1}{r||}{$\infty$}\\
\hline
\mc{1}{|c||}{4.20}	& \mc{1}{r|}{-0.9315}		& \mc{1}{r||}{-0.9288}		& \mc{1}{r|}{-0.0896}	& \mc{1}{r|}{1.4217}	& \mc{1}{r|}{0.9021}		& \mc{1}{r|}{0.000000}	& \mc{1}{r|}{0.000000}		& \mc{1}{r|}{$ $} & \mc{1}{r||}{$ $}\\
\hline
\mc{1}{|c||}{4.50}	& \mc{1}{r|}{-0.9544}		& \mc{1}{r||}{-0.9517}		& \mc{1}{r|}{-0.0705}	& \mc{1}{r|}{1.3742}	& \mc{1}{r|}{0.9263}		& \mc{1}{r|}{0.000000}	& \mc{1}{r|}{0.000000}		& \mc{1}{r|}{$ $} & \mc{1}{r||}{$ $}\\
\mc{1}{|c||}{5.00}	& \mc{1}{r|}{-0.9760}		& \mc{1}{r||}{-0.9732}		& \mc{1}{r|}{-0.0471}	& \mc{1}{r|}{1.3200}	& \mc{1}{r|}{0.9556}		& \mc{1}{r|}{0.000000}	& \mc{1}{r|}{0.000000}		& \mc{1}{r|}{$ $} & \mc{1}{r||}{$ $}\\
\mc{1}{|c||}{7.00}	& \mc{1}{r|}{-0.9939}		& \mc{1}{r||}{$ $}		& \mc{1}{r|}{-0.0082}	& \mc{1}{r|}{1.2504}	& \mc{1}{r|}{0.9972}		& \mc{1}{r|}{0.000000}	& \mc{1}{r|}{0.000000}		& \mc{1}{r|}{$ $} & \mc{1}{r||}{$ $}\\
\hline
\mc{1}{|c||}{$\infty$}	& \mc{1}{r|}{-1.0000}		& \mc{1}{r||}{-1.0000}		& \mc{1}{r|}{0.0000}	& \mc{1}{r|}{1.2500}	& \mc{1}{r|}{1.0000}		& \mc{1}{r|}{0.000000}	& \mc{1}{r|}{0.000000}		& \mc{1}{r|}{$ $} & \mc{1}{r||}{$ $}\\
\hline
\hline
\end{tabular}}
\end{table*}

For the sake of completeness, we also provide in Tab.~\ref{tab:01} the numerical values of the selected quantities vs. $R$. Note that we compared the values of $E_G$ in SGA and GA.
The slightly higher values of $E_G$ obtained in the present (SGA) case should not be surprising for $R$ away from $R_c$. It is rewarding that they are lower as $R_c$ is approaching,
since then the results are more realistic, i.e. the tight-binding approximation works better. The last column describes the effective mass enhancement ($q_\sigma ^{-1}=\frac{m*}{m_B}$)
due to the interparticle correlations. One should note that the bare band value $m_B$ is also dependent on $R$. Additionally, for the half-filled band case (and only then)
the mass enhancement is not explicitly spin-direction ($\sigma = \pm 1$) dependent. The scaling with $R$ of the selected just discussed quantities is carried out next.

\subsection{Scaling laws of physical parameters}
\label{ssec:scal}

\begin{figure}[H]
\centering
\includegraphics[width=0.99\linewidth]{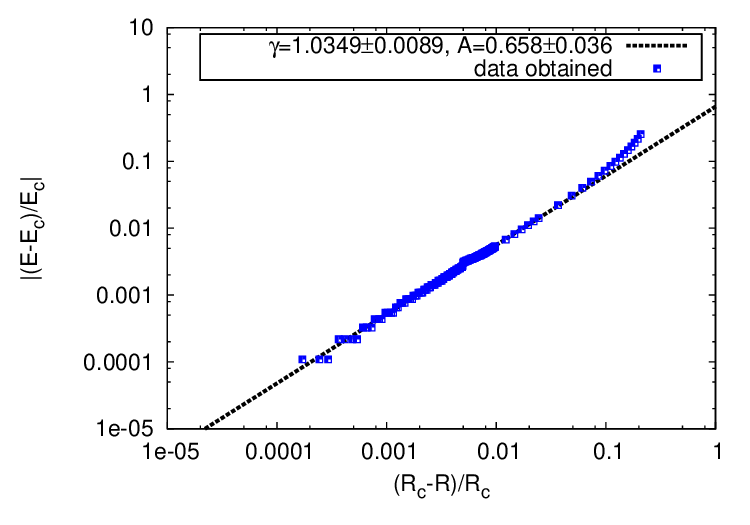}
\caption{Scaling of the relative ground state energy near $R=R_c$. For detailed explanation see main text.
}
\label{fig:log_E}
\end{figure}

The question arises whether the scaling properties proposed earlier \cite{Spalek} of selected physical quantities near the quantum critical point 
(located at $R=R_c$), where $\chi$ is divergent, take place also within the present SGA approach.
To address this question, we have plotted in Fig. \ref{fig:log_E} the relative ground state energy per site $| (E_G(R)-E_G(R_c))/{E_G(R_c)} | \equiv 
| (E-E_C)/{E_C} |$ vs. relative interatomic distance $(R_c-R)/{R_c}$ for $R\rightarrow R_c - 0$. As before, we observe an almost linear scaling
$| (E-E_C)/{E_C} | \\= A ( (R_c - R)/{R_c} ) ^\gamma$, since $\gamma=1.035 \pm 0.009$ and additionally, $A=0.66 \pm 0.04$. One should note that this type
of scaling for the ground state energy near QCP is very different from the corresponding behavior in the atomic limit $R \gg R_c$, where the corresponding energy per site is roughly
dominated by the Coulomb contribution, so $E_G \sim R^{-1}$. This type of scaling means that the first derivative is nonzero and constant for $R \rightarrow R_c$.
This feature will be taken upon and interpreted in Section \ref{ssec:Hmet}.

\begin{figure}[H]
\centering
\includegraphics[width=0.99\linewidth]{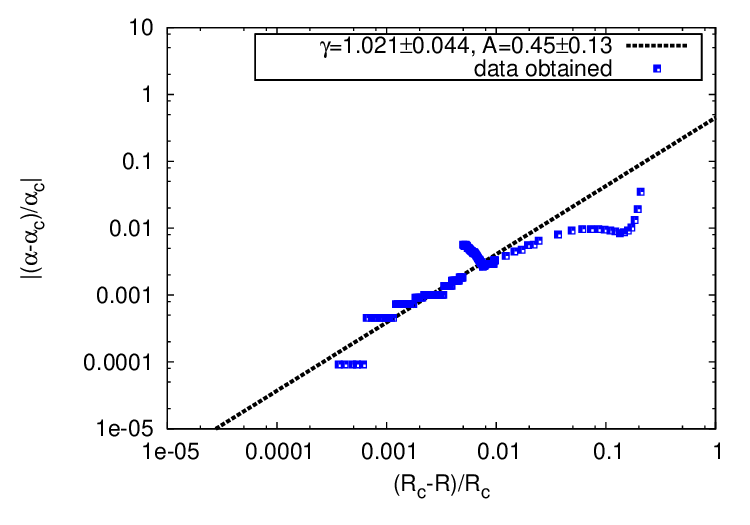}
\caption{Scaling of the inverse orbital size $\alpha$ near $R=R_c$. For detailed explanation see main text.
}
\label{fig:log_a}
\end{figure}

\begin{figure}[H]
\centering
\includegraphics[width=0.99\linewidth]{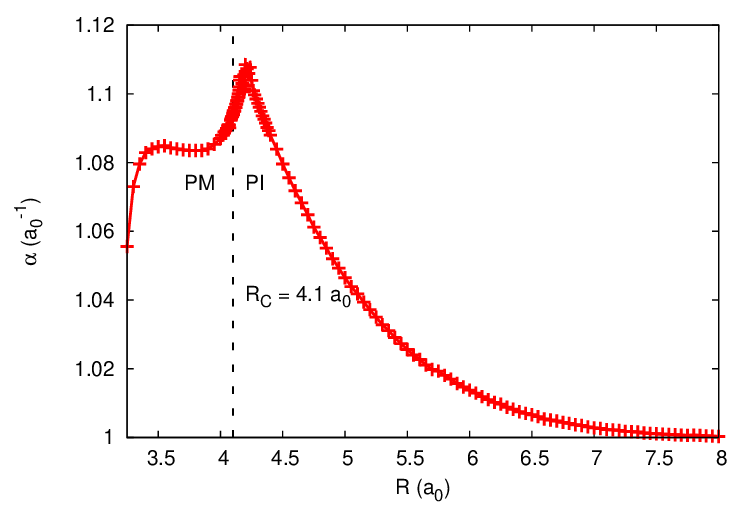}
\caption{Inverse atomic-orbital size vs. $R$. The critical behavior near $R=R_c=4.1 a_0$. The dependence is practically independent of the applied field. Note that the cusp has its
maximum slightly above $R_c$.
}
\label{fig:alpha}
\end{figure}

In Fig. \ref{fig:log_a} we plot the relative inverse renormalized atomic-orbital size $| (\alpha(R)-\alpha(R_c)) / {\alpha(R_c)} | \equiv
| (\alpha-\alpha_c)/ \- {\alpha_c} |$ vs. $(R_c-R)/ \- {R_c}$. Again, we have an almost linear scaling, though the behavior is not as regular as before \cite{Spalek}.
It is thus more legible to display the actual behavior of the renormalized orbital size $\alpha(R)$ vs. $R$, as shown in Fig. \ref{fig:alpha}. The quantum critical behavior
corresponds to the cusp-like feature around $R_c$, where we observe a clear discontinuity of ${d \alpha}/{d R}$ at $R=R_c$. The most important feature of this figure is the fact
that we observe this type of quantum critical behavior for the renormalized inverse size of the single-particle wave function. We are not aware of any other result of that type
appearing in the literature, apart from the analogical result obtained by us earlier within the GA \cite{Spalek}. Note also, as already said earlier, this type of critical behavior
will translate into the same type of behavior for the renormalized Wannier functions. It is important
to emphasize that this quantum critical behavior of the principal characteristic ($\alpha$) of the renormalized wave function is independent of the fact whether the actual two-particle
characteristic - $d^2$ vanishes continuously (as in GA) or discontinuously (as in SGA), as well as whether the states are paramagnetic ($PM$, $PI$) or have a small ''parasitic``
magnetic moment. This analysis should certainly be extended and carried out for the antiferromagnetic ($AFM$, $AFI$) cases, but that requires a separate treatment, as it will
certain a large number of additional variational parameters. This is certainly planned as the next stage of work.

\begin{figure}[H]
\centering
 \includegraphics[width=0.99\linewidth]{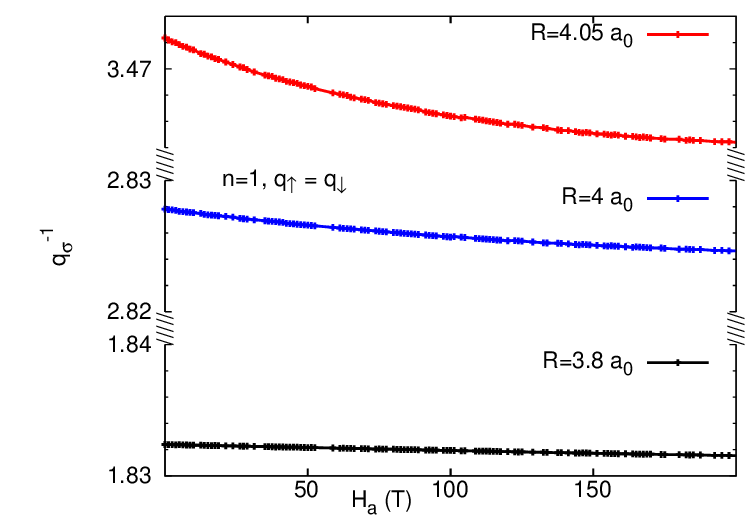}
\caption{Effective mass enhancement as a function of magnetic field for the situation with one electron per atom. In this case it is spin direction independent and dependent
only on the interatomic distance. Note that this enhancement does not include the change of the bare mass which increases also with the increasing $R$.}
\label{fig:gutzmass}
\end{figure}

Finally, we would like to discuss at this point two important physical characteristics of our correlated single narrow-band system. First of them is exhibited
in Fig. \ref{fig:gutzmass} and shows the effective mass enhancement $q_\sigma ^{-1}$ vs. $H_a$ for selected values of $R$ close to $R_c$, but on the metallic side. We observe
the two important features. First, the quasiparticle mass is spin-independent in this half-filled band case, unlike in the non-half-filled case \cite{Spalek2,Spalek15}. Also, the mass
enhancement due to the correlations increases in SGA relatively slowly compared to that in the parametrized GA approach \cite{Spalek3} as $R\rightarrow R_c - 0$. Second, this slow
increase of $q^{-1}$ when confronted with the rather fast dependence of the susceptibility (cf. Fig.~\ref{fig:sus}, Tab. \ref{tab:01}) means that the magnetic contribution to the renormalized
susceptibility (corresponding to the ''renormalized Stoner part``, cf. \cite{Brinkman,Spalek3}) has an essential contribution to the susceptibility. Also, as mentioned earlier, there
may be an essential contribution due to the bare bandwidth $W$ narrowing with the increasing $R$. In effect, the resultant behavior of quantities such as $\chi$ or the linear
specific-heat coefficient $\gamma = \gamma_0 q^{-1}$ in the strong-correlation limit (here $U \approx W$) is much more subtle than it was discussed within the pure parametrized model
(GA) picture, where the renormalization of the Wannier functions, as well as the correlation fields $\lambda_m$ and $\lambda_n$, are absent. Simply put, our approach goes beyond
the parametrized-model approach as it provides the correlated system evolution as a function of experimentally accessible parameter - the lattice parameter.

\begin{figure}[H]
\centering
\includegraphics[width=0.99\linewidth]{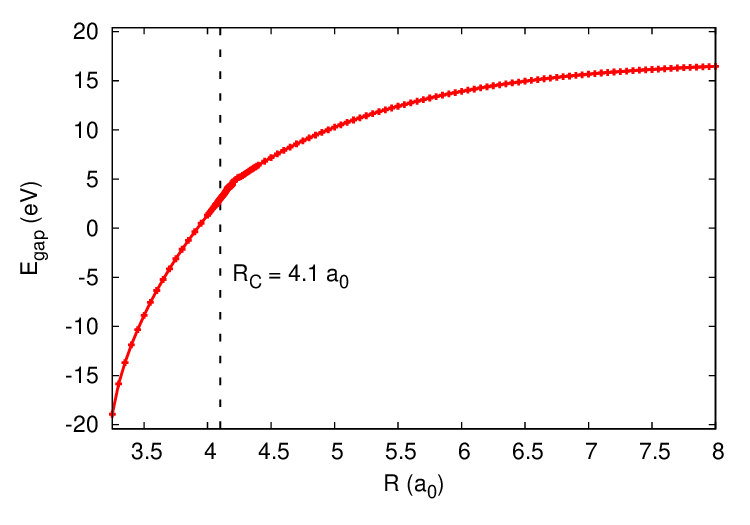}
\caption{Hubbard gap evolution $E_{gap} = U-W$ as a function of interatomic distance. Note a continuous dependence near $R_c$ even though the transition is weakly discontinuous
(cf. Figs.~\ref{fig:met-ins}, \ref{fig:dpow2}). As the gap is increasing rapidly with the increasing $R$ ($R>R_c$), the Mott insulators fir $R>4.5 a_0$ cannot be regarded
as wide-gap semiconductors.}
\label{fig:E_gap}
\end{figure}

Finally, in Fig. \ref{fig:E_gap} we plot the value of the Mott-Hubbard gap $E_{gap}=U-W$ versus $R$. Obviously, in the metallic phase it does not appear (for the sake of completeness,
we show it as having a negative value). Above the transition it reaches relatively fast with the increasing $R>R_c$ the asymptotic value $E_{gap} \sim U$. A weak cusp-like behavior
should be noted just above $R=R_c$. This feature is reminiscent of the cusp-like behavior of the renormalized Wannier functions. It would be interesting to study the pressure dependence
of the gap for the Mott insulator located just above the transition point and determine its scaling with $(R-R_c)/R_c$. It does not scale linearly, as the quantities earlier.

\subsection{Supplement: Application to solid atomic hydrogen: estimated of the critical pressure for metallization}
\label{ssec:Hmet}

\begin{figure}
\centering
\includegraphics[width=0.99\linewidth]{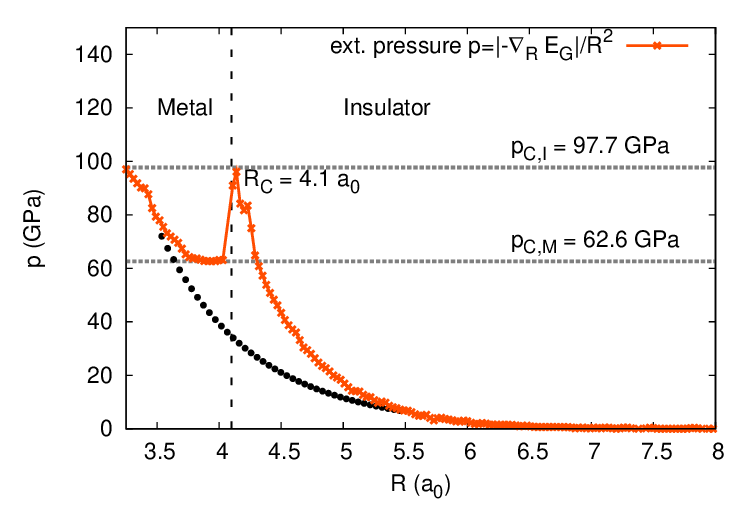}
\caption{External pressure (in $GPa = 10^9 Pa$) one has to exert in order to stabilize the crystal vs. interatomic distance (with cell area $A/N = R^2$). Note two critical values
of pressure: $p_{C,I}=97.7 GPa$ required in the insulating state and $p_{C,M}=62.6 GPa$ in the metallic state. The lattice becomes very rigid as $R\rightarrow R_c$. The dotted
line marks naive extrapolation $M\rightarrow I$.}
\label{fig:pres}
\end{figure}

Our formulation represents a fully microscopic model of the solid atomic hydrogen undergoing a Mott insulator-metal transition. This model differs from the standard treatment
\cite{Shibata}, in which the phase diagram is treated as a function of microscopic parameter $U/W$, since we include here a fully self-consistent procedure of evaluating
the renormalized-by-correlations wave functions. In effect, we can calculate explicitly the critical pressure for metallization. In this approach the external pressure is
the factor stabilizing the system in a particular phase ($M$ or $I$). This task can be carried out by using
the classical definition of pressure $p$ as the force $F$, applied to make the present atomic solid stable, over the area $A$. This force is obtained by differentiating 
$F= \left| -\nabla_R E_G \right|$. The corresponding external pressure we have to exert on the system in order to stabilize the crystal as a function of interatomic distance
(with energy $E_G$ per site and the elementary cell area $A/N = R^2$) is plotted in Fig.~\ref{fig:pres}. Note that the physical meaning
has the critical pressure for metallization as the point $R=R_c$ is the terminal point of stability of \emph{the almost localized Fermi liquid}. It differs from other results \cite{Weir}
as those results concern a molecular ($H_2$) solid; for detailed comment see e.g. \cite{Silvera}. Note that the pressure behavior $p(R)$ for $R\rightarrow R_c$ traces the difference
on the slopes $dE/dR$ for metallic ($R<R_c$) and insulating ($R>R_c$) regimes. Roughly, it shows a critical (divergent) behavior at $R_c$. To determine that more accurately we
require a refined numerical analysis. Nonetheless, the trend is clear and provides a promising starting point for a further analysis.

One must emphasize that this discussion is based on an implicit assumption that the positive ions (protons in this case) form a lattice in both the metallic ($M$) and
insulating ($I$) states. This may turn only a rather crude approximation on the possible role of their zero-point motion in increasing 
the value of the critical pressure. Additionally, a possible transition to solid-liquid plasma as an alternative scenario, should
certainly not be ruled out (for the relevant discussion of that point e.g. \cite{Morales,Tamblyn}).

\section{Outlook}
\label{sec:con}

We have provided ground-state principal characteristics of the correlated narrow-band system involving first-principle variational calculations of the renormalized
Wannier functions of 1s-type in the correlated state described by the Statistically Consistent Gutzwiller Approximation (SGA). Our model is fully microscopic as it contains
an explicit calculations of microscopic parameters of the extended Hubbard model. In this manner, we can track the correlated Fermi-liquid evolution with the increasing lattice
parameter $R$ into the Mott-Hubbard insulator. As an extra bonus, we obtain an estimate of the critical pressure for the metallization of solid atomic hydrogen which is roughly
equal to $100 GPa$. Obviously, the solid atomic hydrogen is a molecular crystal up to much higher pressures \cite{Shibata,Weir}, but it may be possible that the states of this type
can be synthesized in the optical-lattice experiments.

We would like to emphasize that our model calculations contain two attractive features from a general methodological point of view of the computational solid state
physics. First, they do not include twice the Coulomb interactions among electrons, as it is the case for LDA+U or LDA+DMFT methods. Those latter methods have their own merits
as they can be (and are) applied to complex real materials. But the correctness of the present approach is still worth mentioning, even when applied so far to the mode system
only. Second, within the present approach the calculation of the renormalized characteristics of the single-particle wave function and of the interelectronic
correlations, are both treated on the same footing. This is the crucial
feature of the Mott-Hubbard systems, for which the single-particle energy (e.g. the renormalized band energy) is of the same magnitude as the interparticle (Coulomb in this
case) interaction. The approach of the present type should be extended to the multiband situation to discuss the realistic strongly correlated materials (the magnetic
oxides such as $V_2O_3$) evolving, when varying the lattice parameter (i.e., applying the pressure), and not only as a function of the microscopic parameters such as $U/W$,
as they vary very rapidly in the vicinity of the Mott-Hubbard transition. However, this program execution may represent a long road ahead.

One has to mention that the present approach requires some other basic extensions. First of all, as any Gutzwiller-type approach, it does not include explicitly the intersite
kinetic exchange appearing deeply in the Mott insulating state (for $R$ substantially larger than $R_c$). The trace of this interaction is coded in the correlation fields
$\lambda_m$, as has been shown in a related context before \cite{Korbel}. Second, the case with partial band filling ($n<1$) should be also treated. Third and most importantly,
the antiferromagnetic state should be included in the analysis. This is our plan for the near future. However, we believe that this simple analysis shows up in a clear
manner the quantum critical behavior of the renormalized-by-correlations single-particle wave function, not obscured by the complicated magnetic structure (i.e., that
with a staggered magnetic moment). Fourth and finally, the approach, even in the present single-band case, should be extended to the close packed lattices such as \emph{fcc}
and \emph{hcp}, as this are the typical structures for metals. But then, one has to include at least first two hopping integrals (between the nearest and the next-nearest
neighbors). Such treatment is certainly tractable. Nonetheless, we believe that our first estimate of the critical pressure for the metallization of solid atomic hydrogen,
carried out for the \emph{sc} structure, provides a promising starting point.

Finally, one can extend our analysis to nonzero temperatures in a straightforward manner, starting from the free-energy functional \eqref{eq:SGA}. This should be
carried out separately and the results can be compared with GA results providing the first-order line ending in a classical critical point, as well as the reentrant
metallic behavior \cite{Spalek4,*Spalek5,Spalek2,Spalek15}.

\section{Acknowledgments}
\label{sec:ack}
\begin{acknowledgement}

Discussions with Jan Kaczmarczyk are greatly appreciated. The work was carried out partly within the Project TEAM awarded to our group by the Foundation for Polish
Science (FNP) for the years 2011-2014 (APK and JS).
We also acknowledge partial support by the special Grant MAESTRO from the National Science Center (NCN) for the years 2012-2017.
\end{acknowledgement}

\bibliographystyle{epj}
\bibliography{bibliography}
\end{document}